# Raman and x-ray diffraction investigations on $BaMoO_4$ under high pressures

Vinod Panchal, Nandini Garg and Surinder M. Sharma*

Synchrotron Radiation Section

Bhabha Atomic Research Centre, Mumbai 400085, India

**Abstract**

X-ray diffraction and Raman scattering studies on the scheelite structured barium molybdate show that, at ~ 5.8 GPa, it undergoes a first order phase transition to the fergusonite structure ($I2/a$, $Z = 4$) - as also observed in iso-structural barium tungstate. At still higher pressures, barium molybdate transforms to another phase between ~ 7.2 – 9.5 GPa. On release of pressure from 15.8 GPa, the initial phase is recovered implying that the observed structural modifications are reversible.

*Corresponding author : Fax: +91-22-25505151
                        Tel.: +91-22-25595476
Email address: smsharma@magnum.barc.ernet.in (Surinder M Sharma)

# 1. Introduction

Scheelite structured alkaline earth tungstates and molybdates are technologically important materials as these are used as scintillators [1], laser host materials [2] or cryogenic detectors for dark matter [3]. These tungstates are also being considered for Raman lasers [4]. In these scheelite structured compounds (Space Group $I4_1/a$, Z = 4), oxygen atoms form a distorted simple cubic arrangement, while the two cations in this structure have fourfold and eightfold coordination with the oxygen atoms. To understand the structural stability under different thermodynamic conditions, many scheelite compounds have been extensively studied [5-12]. In particular, under high pressures, the scheelite structured alkaline earth molybdates and tungstates show several phase transitions [7-12]. Based on the packing efficiency considerations, some of the earlier studies on these compounds suggested that wolframite [8,13] may be a preferred high pressure structure [14]. However, recent high pressure investigations have shown that these compounds may transform to either of the competing monoclinic structures, such as wolframite, [8,13] fergusonite, [7,9,15] or $HgMoO_4$ [12]. For example, barium tungstate [9] has been shown to transform to the fergusonite structure at ~ 7 GPa. Even $CaWO_4$, which was viewed as a strong candidate for the wolframite structure at high pressures [16], has been shown to transform to the fergusonite phase at 11.2 GPa [17]. Recent XANES and x-ray diffraction studies by Errandonea et al [18] have shown that $SrWO_4$ also belongs in the same category i.e. it too transforms to the fergusonite phase at 9.9 GPa. In addition, their ab initio total energy calculations show that for $CaWO_4$ and $SrWO_4$, an orthorhombic Cmca phase is more stable than the fergusonite phase beyond 29 GPa and 21 GPa respectively. However, these calculations also indicate that $LaTaO_4$ structure may be in close competition with the Cmca phase for the second high pressure phase for $SrWO_4$. Recently the high pressure behaviour of another member of this group, $BaMoO_4$, has been investigated with the help of Raman studies up to 8 GPa and it indicates a phase transition at ~ 5.8 GPa [19]. To gain further understanding of the structural chemistry of this family, and also to determine the structure of the high pressure phase in $BaMoO_4$ we have carried out synchrotron based angle dispersive x-ray diffraction experiments. In addition, Raman measurements were also extended up to ~ 13

GPa. We present here our results of these investigations and compare these with the results of other closely related compounds.

**2. Experimental**

High pressure Raman and x-ray diffraction studies were carried out on the powder samples of barium molybdate (Aldrich chemicals (purity 99.9 %)). Raman spectra were recorded using a CCD based single stage spectrograph employing a super-notch filter. 514.5 nm line of the argon ion laser was used as an excitation source. Raman signal was collected in the back scattering geometry, with the laser beam incident upon the sample at an angle of ~ 45° with respect to the axis of collection optics. Several modes lying in between 250 $cm^{-1}$ and 1000 $cm^{-1}$ were recorded as a function of pressure. Reasonably good quality Raman data could be obtained with typical collection times of ~ 100 sec.

Angle dispersive x-ray diffraction measurements were carried out at the XRD1 beamline at Elletra synchrotron source using monochromatic x-rays of wavelength 0.6967 Å, (calibrated using a $LaB_6$ NIST standard). The x-rays were collimated to ~ 100 μm and the two dimensional diffraction rings, collected on a MAR 345 imaging plate, were converted to one dimensional diffraction profiles using the FIT2D software [20]. The cell parameters were determined using Le Bail analysis incorporated in the GSAS software [21]. Using the same software, Rietveld analysis was also carried out for the initial scheelite and the first high pressure phase.

For both the studies, powder sample of $BaMoO_4$ was loaded in a hole of ~ 130 μm diameter of a pre-indented (80 μm) tungsten gasket, which was mounted in a Mao-Bell type of diamond anvil cell. 4:1 methanol-ethanol mixture was used as a pressure transmitting medium. In the Raman experiments ruby R-lines were used for the pressure calibration, [22] whereas for the x-ray diffraction experiments gold was used as a pressure marker. The pressure on the sample in the x-ray diffraction experiments was deduced using the equation of state of gold as given in ref. 23.

## 3. Results and Discussion

### 3.1 Raman Scattering

At ambient conditions, BaMoO$_4$ exists in the scheelite structure (space group I4$_1$/a and point group C$^6_{4h}$) with two formula units per primitive cell (four formula units in the conventional body centered tetragonal unit cell). The group theoretical analysis predicts that the optical phonons belong to the following irreducible representations of the point group C$^6_{4h}$, 3A$_g$ + 5B$_g$ + 5E$_g$ + 4A$_u$ + 3B$_u$ + 4E$_u$. Of these, the A$_g$, B$_g$, and E$_g$ modes are Raman active and A$_u$, B$_u$, and E$_u$ are infrared active. Raman active internal modes due to MoO$_4$ tetrahedra are identified as $\nu_1$ (A$_g$), $\nu_3$ (B$_g$ and E$_g$) (stretching modes), and $\nu_2$ (A$_g$ and B$_g$), $\nu_4$ (B$_g$ and E$_g$) (bending modes) [12]. At ambient conditions, we have retained the assignments given by Jayaraman et al, [12] which were based on the earlier work of Liegeois-Duyckaerts and Tarte [24].

Raman spectra of BaMoO$_4$ at a few representative pressures are shown in Fig. 1. The observed Raman modes above 250 cm$^{-1}$ are known to be the internal modes of MoO$_4$ tetrahedra. Figure 2 shows that, in the scheelite phase, all the observed modes display a gradual stiffening with pressure. At ~ 6 GPa, some new modes appear at ~ 309 cm$^{-1}$ ($\omega_1$), 806 cm-1 ($\omega_5$), 835 cm$^{-1}$ ($\omega_8$) and 866 cm$^{-1}$ ($\omega_{10}$). On further increase of pressure to 6.8 GPa, several of these changes in the Raman spectra became more clearly visible. Moreover, around this pressure, the accidental degeneracy of $\nu_2$ (A$_g$+B$_g$) modes was removed, as has also been observed in iso-structural SrWO$_4$ and CaWO$_4$ [25]. In addition, Raman modes show a change in slope (d$\omega$/dP) around this pressure. These observed changes in the Raman spectra indicate a phase transition, probably to a lower symmetry phase (Phase II). These results, before and across ~ 6 GPa, are similar to those obtained recently by Christofilos et al [19]. It is also interesting to note that the pressure of this transformation is quite close to the transformation pressure of ~ 7 GPa observed in iso-structural BaWO$_4$ [12].

The slope ($d\omega_i/dp$) and the corresponding mode-Grüneisen parameters of various Raman modes are listed in table 1. For mode-Grüneisen parameters, we have used the bulk modulii obtained from our x-ray diffraction studies (presented in the next section). The slopes of most of the internal modes in the scheelite phase have comparable values. Comparison of the corresponding Grüneisen parameters for $BaMoO_4$, $BaWO_4$ and $CaMoO_4$ shows that these are also qualitatively similar. These results suggest that the pressure induced changes in the force constants of these scheelite structured compounds are similar, particularly for the higher frequency modes.

On increasing the pressure further, at ~ 7.7 GPa we observed broadening of several Raman peaks along with the appearance of a new mode at 820 cm$^{-1}$ (marked with an arrow in Fig. 1). At still higher pressures, another mode also starts developing at 450 cm$^{-1}$ accompanied by a change of slope of the Raman modes a ~ 9 GPa. At ~ 9.5 GPa an additional new Raman mode was also observed at ~ 675 cm$^{-1}$. In addition, the intensity of the mode at 806 cm$^{-1}$ also increased significantly. No more changes were observed in the Raman spectra till ~12.7 GPa. These observed changes in the Raman spectra may not be attributable to the non-hydrostatic pressures arising due to the freezing of the liquid pressure medium, as even in our x-ray diffraction studies (to be discussed in the next section) we observed new diffraction peaks around this pressure. Due to the continuation of all the modes from 7.7 to 9.5 GPa and from the emergence of new modes, we conclude that another phase change which is initiated between 7-7.7 GPa is completed by ~ 9.5 GPa. Moreover, the appearance of a new mode at 675 cm$^{-1}$ i.e. in the gap of ~ 400-780 cm$^{-1}$ of the scheelite and the first high pressure phase (phase –II), suggest that the structure of the new phase may be quite different from the preceding phases. It should be noted that similar second phase transition was also observed in several other scheelite structured compounds like $YLiF_4$ and $BaWO_4$ at 17 GPa, and 14 GPa respectively [7,9]. Recent first principles total energy calculations by Errandonea et al [18] also predicted that $CaWO_4$ may transform to a second high pressure phase at ~ 29 GPa and $SrWO_4$ may do so at ~ 21 GPa. Their results suggest that in the new structure (Cmca) counter-cation (Ca and Sr) would be coordinated to six oxygen atoms. Such a coordination change in $BaMoO_4$ may be consistent with the appearance of the new Raman modes at 675 cm$^{-1}$.

### 3.2 X-ray diffraction

To determine the structure of the new phases observed in the Raman spectra, x-ray diffraction studies were carried out. X-ray diffraction profiles of barium molybdate at a few pressures are shown in Fig. 3. The diffraction peaks of gold and tungsten are marked as Au (hkl) and W (hkl) respectively. Before we present the analysis, we would list the observed changes in the diffraction pattern under pressure. The diffraction pattern evolves smoothly up to 5 GPa. At ~ 5.8 GPa most of the diffraction peaks showed broadening and weak shoulders developed at 2θ ~ 15.2° and 20° (marked with arrows in Fig. 3). However, at 5.8 GPa, the diffraction pattern is still indexable in the scheelite phase. At ~ 7.2 GPa, apart from the broadening of the diffraction peaks, several new diffraction peaks were also observed and the shoulders that appeared at 5.8 GPa also developed into clearly identifiable diffraction peaks. The splitting of the (200) and intensity redistribution of (312) and (224) diffraction peaks (marked with * in Fig. 3) is also easily identifiable. These results indicate that beyond 5.8 GPa $BaMoO_4$ transforms to phase II as also implied by the changes in the Raman spectra at ~ 6 GPa. On further increase of pressure to ~ 8.6 GPa, a new diffraction peak appeared at ~ 2θ = 5.8° and weak shoulders started developing at 2θ ~ 14.4° and 20.4°. Apart from these changes the rest of the diffraction pattern was similar to that of phase II. At 9.5 GPa the intensity of the diffraction peak at ~ 2θ = 5.8° further increased and this is also accompanied by more changes in the diffraction pattern. Detailed analysis of the evolution of the diffraction pattern (given in next paragraph) shows that phase II begins to transform to another phase at ~ 7.2 GPa and this transformation is completed by ~ 9.5 GPa. The appearance of a diffraction peak at lower 2θ value indicates unit cell enlargement. This phase continues to exist up to the highest pressure of this study i.e., 15.8 GPa. On release of pressure the initial phase is fully recovered as observed in other molybdates and tungstates of this family of compounds [9,10,26]. However, as shown in Fig. 3, the diffraction peaks retain some broadening on release of pressure. This may be due to the residual strains in the polycrystalline samples, as this pattern was recorded immediately after release of pressure from 15.8 GPa.

At ambient conditions 26 diffraction peaks of the scheelite phase (phase I) are identifiable. Some of the hkl indices for the scheelite phase are marked in Fig. 3. The lattice parameters of the scheelite phase are found to be a = 5.58 ± 0.01 Å and c = 12.82 ± 0.01 Å, which are in very good agreement with the published lattice parameters, viz., a = 5.58 ± 0.01 Å and c = 12.82 ± 0.01 Å (ICDD card no 08-0455). Rietveld analysis of the diffraction pattern at ambient conditions shows that the atomic positional parameters (shown in table II) are in good agreement with the earlier reported values.

Though the Rietveld analysis of diffraction pattern at ~ 5.8 GPa suggests that it could be refined with the initial tetragonal phase, somewhat poorer nature of the fit implies that the structure at this pressure might have deviated from the initial structure. However, the analysis of the diffraction pattern at 7.2 GPa indicates that a phase transformation has taken place, probably to a lower symmetry phase (phase II), in agreement with our Raman results. Le Bail analysis of the diffraction pattern at this pressure showed that it can be fitted to a monoclinic space group with I2/a symmetry (i.e. fergusonite - a distorted scheelite structure) except for a weak diffraction peak at 2θ ~ 11.8° [27]. For Rietveld analysis of phase (II), the initial coordinates were obtained by transforming the coordinates of the scheelite phase at 5.8 GPa to the coordinates of the I2/a space group using the group-subgroup relations as incorporated in the powdcell software [28]. A typical fit to the data at this pressure is shown in Fig. 4 and the parameters of goodness of fit are $R_{wp}$ = 8.7 %, $R_p$ = 6.6 %. Therefore, it may be reasonable to accept that this phase has a fergusonite structure, similar to the high pressure phase observed in $BaWO_4$ and $CaWO_4$ [9,17]. The structure of this phase is related to the initial scheelite phase through a correlated displacive atomic motion. For this, the lattices of alternate layers of the scheelite structure need shift in opposite directions along the c- axis, forming pairs with the neighbor layers. The a- axis of the monoclinic cell ($a_m$) lies along the a-axis of the tetragonal cell ($a_t$), but $b_m$ lies along $c_t$ and $c_m$ is along the $b_t$ [9].

The cell constants and the coordinates of the new phase have been listed in table II. The Mo-O and Ba-O bond distances for the tetragonal and monoclinic phases are also given in this table. Deduced Mo-O and Ba-O bond lengths in the monoclinic phase imply that MoO$_4$ tetrahedra as well as BaO$_8$ octahedra are distorted in this phase. Moreover reduced bond lengths at ~ 7 GPa imply that in barium molybdate, the phase transformation to the denser phase affects both – the tetrahedra as well as octahedra, unlike in the scheelite phase as discussed later .

The variations of the cell constants and volume (V/V$_0$) with pressure are shown in Fig. 5 a, b, c and d respectively. Across the phase change to the fergusonite phase, the lattice parameters a and b differ from those of the scheelite structure only marginally. However, we observe a decrease of 3.4 % in the **b$_t$**. It is also seen that the **c$_t$/a$_t$** ratio of the tetragonal phase decreases and the **b$_m$/c$_m$** and **b$_m$/a$_m$** ratio of the fergusonite phase increase and decrease respectively, a feature similar to the observations in barium tungstate [9] and calcium tungstate [17]. It should also be noted that in the scheelite fluorites and rare earth niobates and tantalates, the c/a ratio of the scheelite phase is known to increase but the respective ratios of the unit cell constants in the fergusonite phase have similar trends as alkaline tungstates and molybdates. These differences may be due to the trivalent cation of the MX$_4$ (M=Y, Ta, Nb, Lu) tetrahedra in these compounds.

Fig. 5d shows that V/V$_0$ drops by ~ 3.6 % across the phase transition to the fegusonite phase, which is comparable to that observed in barium tungstate. In other scheelites like YLiF$_4$, even though the experimental results show a negligible volume change across scheelite to fergusonite phase transition, electronic energy calculations suggest volume collapse of ~ 0.5 % across the phase transition. In SrWO$_4$ and CaWO$_4$, Errandonea et al. also did not observe any detectable volume discontinuity at the scheelite to fergusonite phase transition.

The P-V data of barium molybdate in the tetragonal phase, fitted to the second order Brich- Murnaghan equation of state gives the bulk modulus to be 56 GPa with K′ = 4. Corresponding values for BaWO$_4$ are 57 GPa K′ = 3.5. These results indicate that in the scheelite phase, the compression of these compounds is similar. These results are also consistent with proposed systematics of Errandonea et al, [18] which show that polyhedral bulk modulus of oxides is approximately given as, K$_p$ (in GPa) = 610 Z$_i$/d$^3$, where Z$_i$ is the cationic formal charge and d is the mean cation-O distance. For BaMoO$_4$,

$Z_i = 2$ and Ba-O ~ 2.73 Å, giving tetrahedral bulk modulus of ~ 60 GPa, close to our observed values. Therefore we conclude, as in other scheelite compounds, [9,29] the compressibility in the scheelite phase is primarily due to the larger and softer Ba-O bonds rather than W-O tetrahedral bonds.

To evaluate the structure of the second high pressure phase (phase III), we have carried out the analysis with several competing monoclinic structures such as $NiWO_4$ wolframite structure (P2/c, Z = 2), $LaTaO_4$ structure ($P2_1/c$, Z = 4) and also the recently proposed Cmca structure (Z = 8). For this the transformed unit cell dimensions and the atomic coordinates were generated for $P2_1/c$, P2/c and Cmca space groups with the help of the PCW software [28]. Thus generated Le Bail fits do not display an unambiguous preference between $P2_1/c$ ($LaTO_4$ type: a = 7.76 Å, b = 13.73 Å, c = 5.49 Å and γ = 100.68°) and Cmca (a = 6.30 Å, b = 13.44 Å and c = 3.17 Å).

## 4. Conclusions

To conclude, our x-ray diffraction and Raman studies on barium molybdate show that it transforms to the fergusonite phase ~ 5.8 GPa, as also observed in iso-structural barium tungstate and calcium tungstate. Beyond 7.2 GPa, the fergusonite phase transforms to another phase which could have either $LaTaO_4$ ($P2_1/c$) or Cmca structure. On release of pressure from 15.8 GPa, this compound reverts back to the initial scheelite phase. Our results suggest that, in spite of different alkaline earth cations in the scheelite structured alkaline earth tungstates and molybdates, these may favorably transform to the fergusonite structure before transforming to a different lower symmetry phase.

the onset of the second high pressure phase transition is at somewhat lower pressures i.e. ~ 7.2 GPa

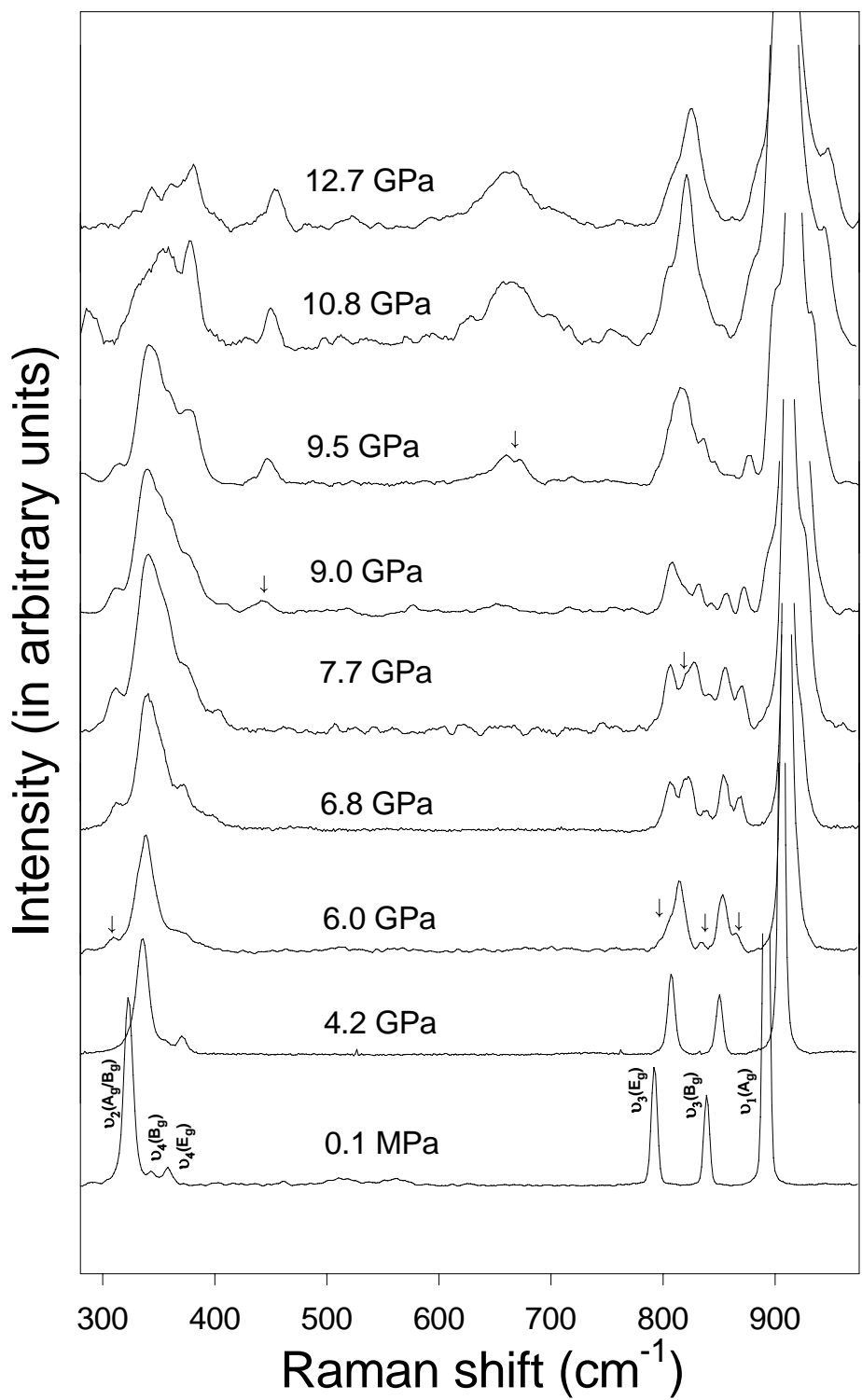

Figure 1. Background subtracted Raman spectra of $BaMoO_4$ at various pressures. The arrows indicate the new Raman modes.

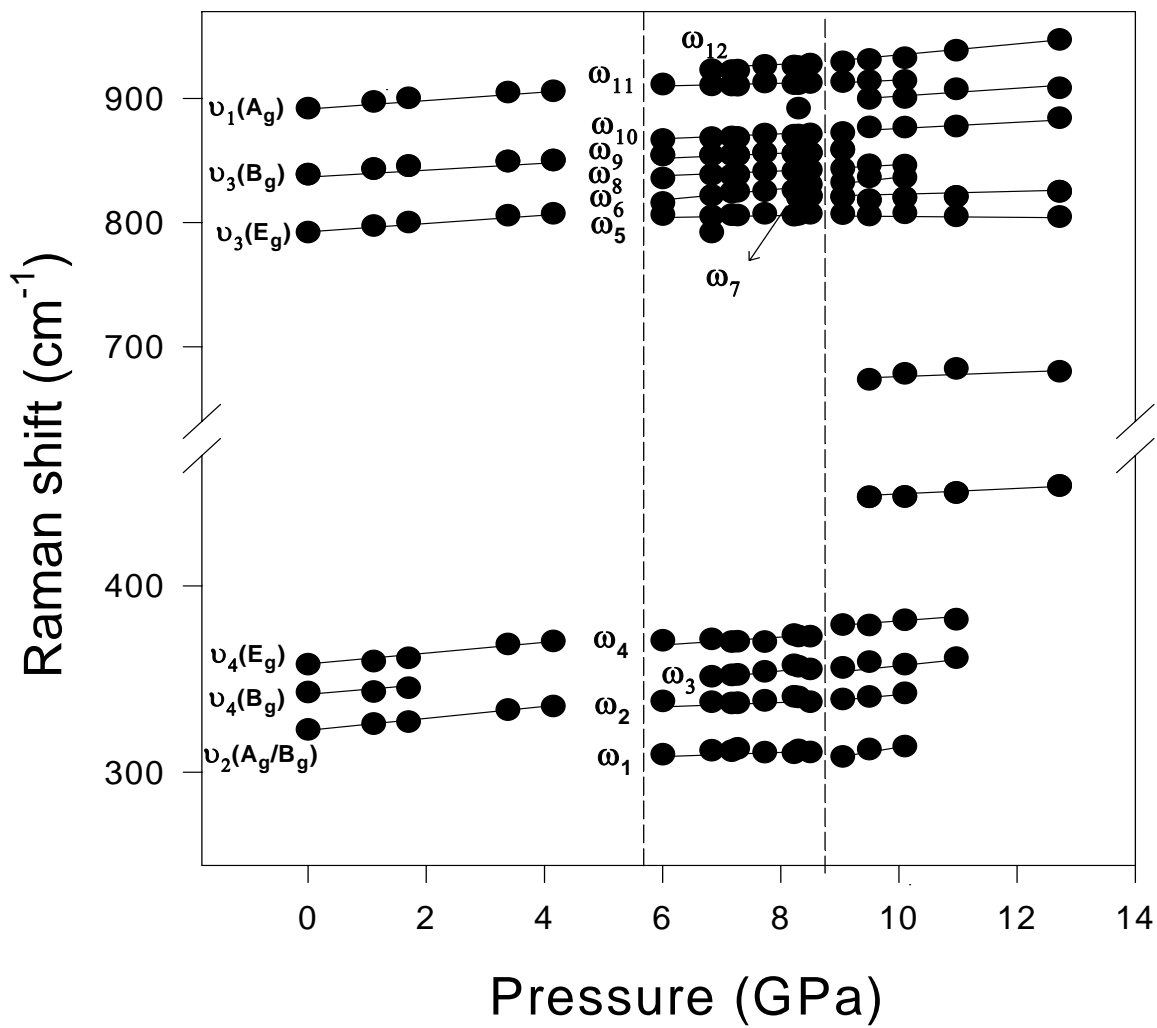

Figure 2. Pressure dependence of Raman modes of barium molybdate. $\nu_1 - \nu_4$ represent the Raman modes of the scheelite phase and $\omega_1$ to $\omega_{12}$ are the Raman modes of the new high pressure phase (fergusonite). The solid lines are drawn as a guide to the eye. The vertical dashed lines indicate the pressures beyond which the phase transitions to the new phases were observed.

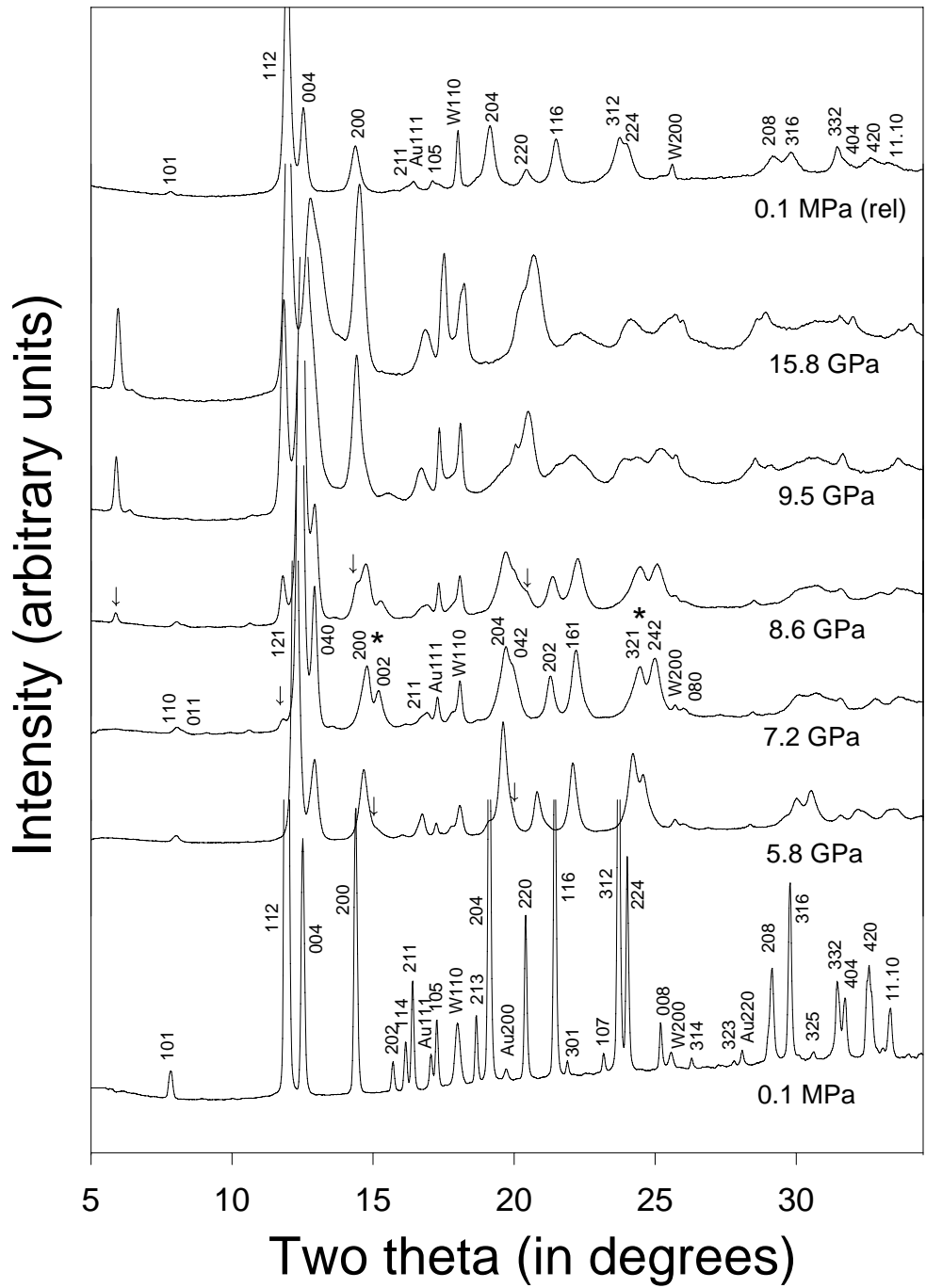

Figure 3. X-ray powder diffraction patterns of BaMoO$_4$ at various pressures. Au (hkl) and W (hkl) are the diffraction peaks from gold and tungsten respectively. The initial and first high pressure phase have been indexed with the scheelite and the fergusonite structure respectively.

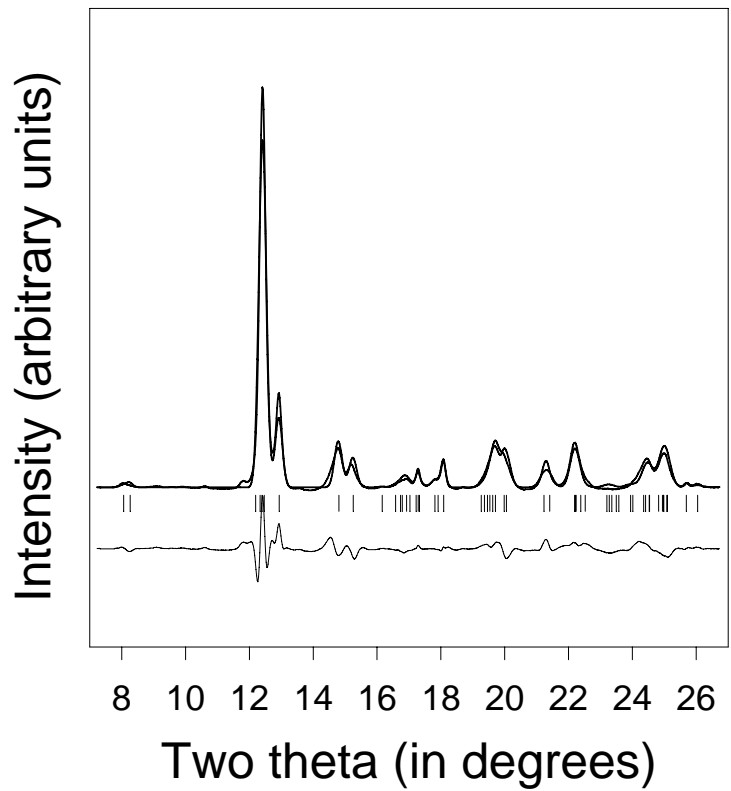

Figure 4. Observed, calculated and difference x-ray diffraction patterns for the fergusonite (I2/a, Z = 4) phase of $BaMoO_4$ at 7.2 GPa. Bars indicate the expected position of diffraction peaks.

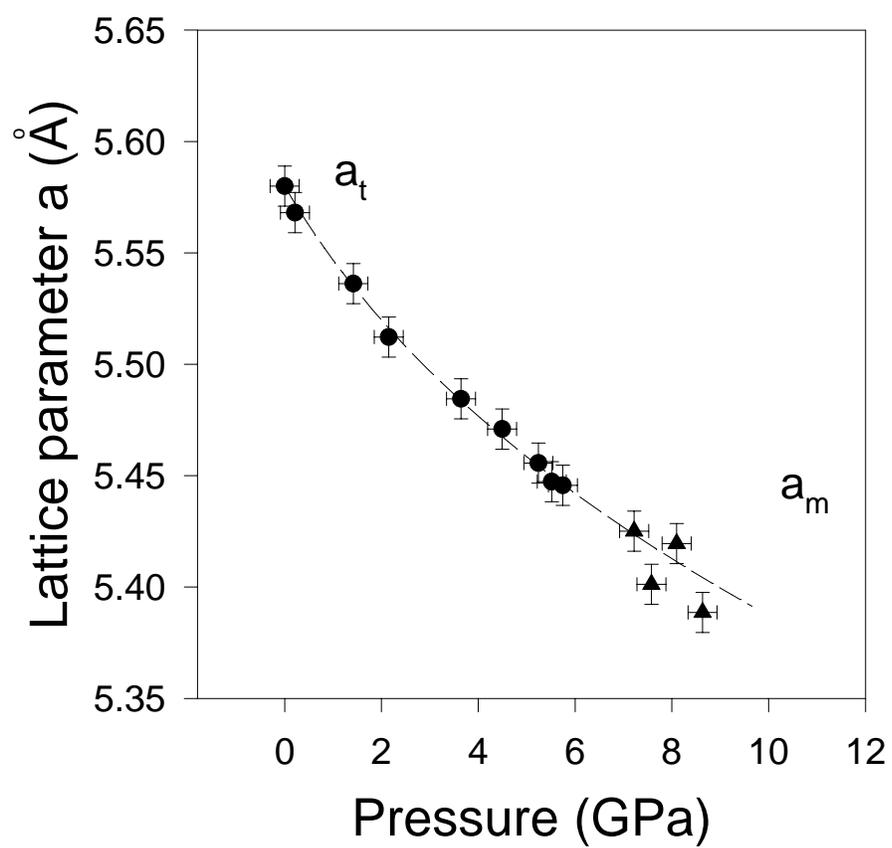

Figure 5(a). Lattice parameter **a** for the scheelite phase (●) and for the fergusonite phase (▲).

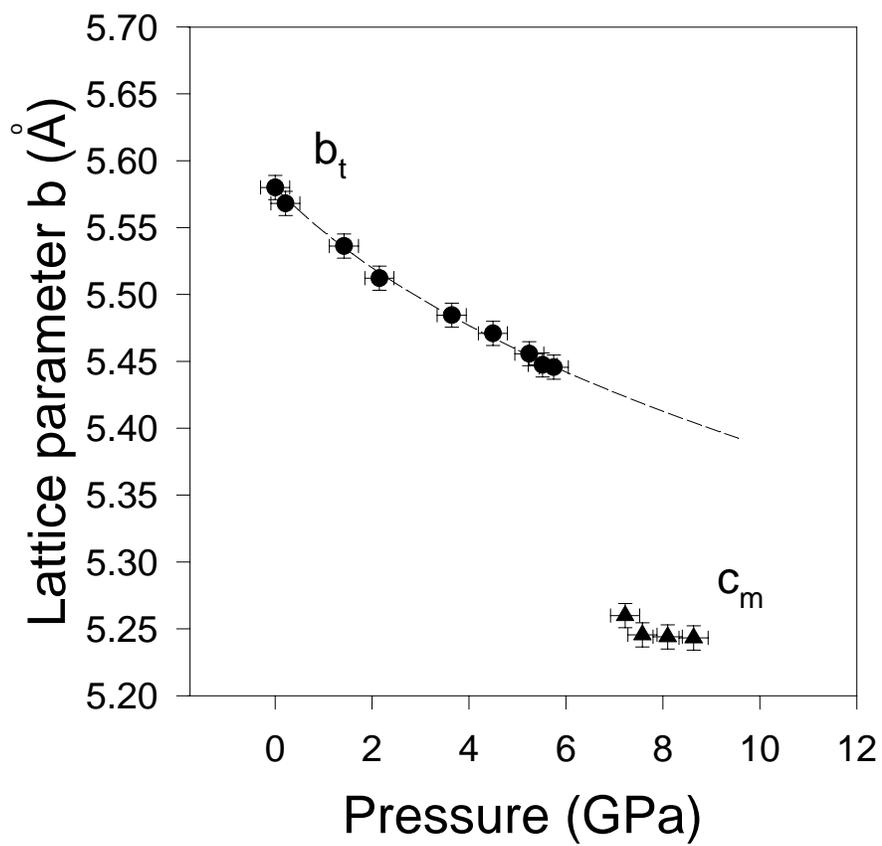

(b) Lattice parameter **b** for the scheelite phase (●) and **c** for the fergusonite phase (▲)

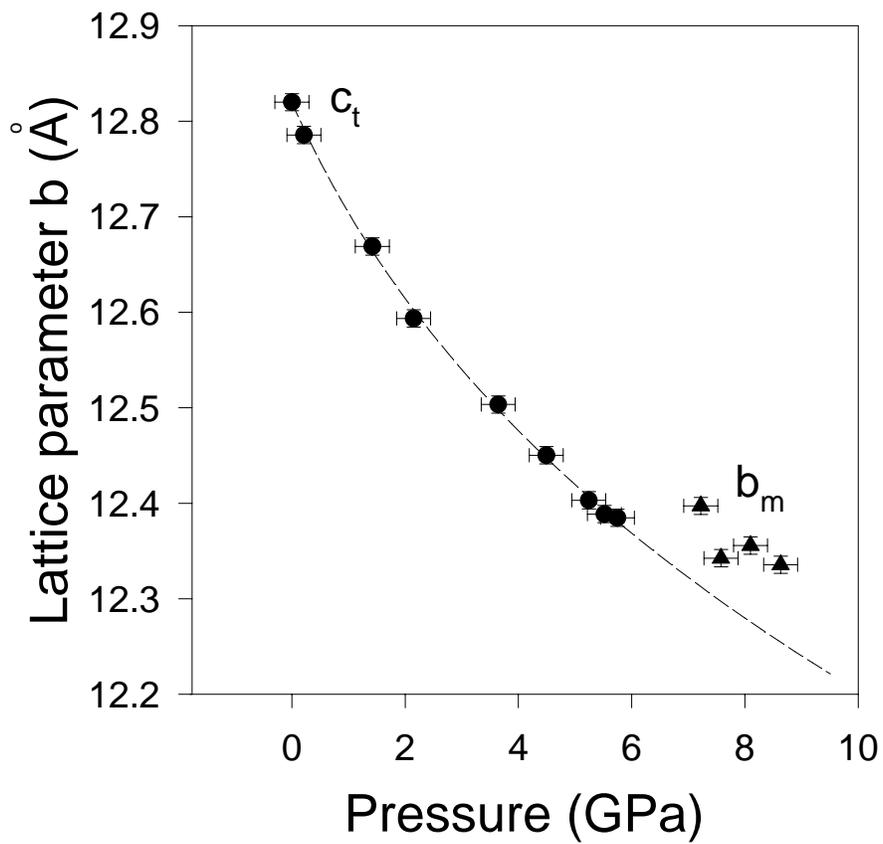

(c) Lattice parameter **c** for the scheelite phase (●) and **b** for the fergusonite phase (▲)

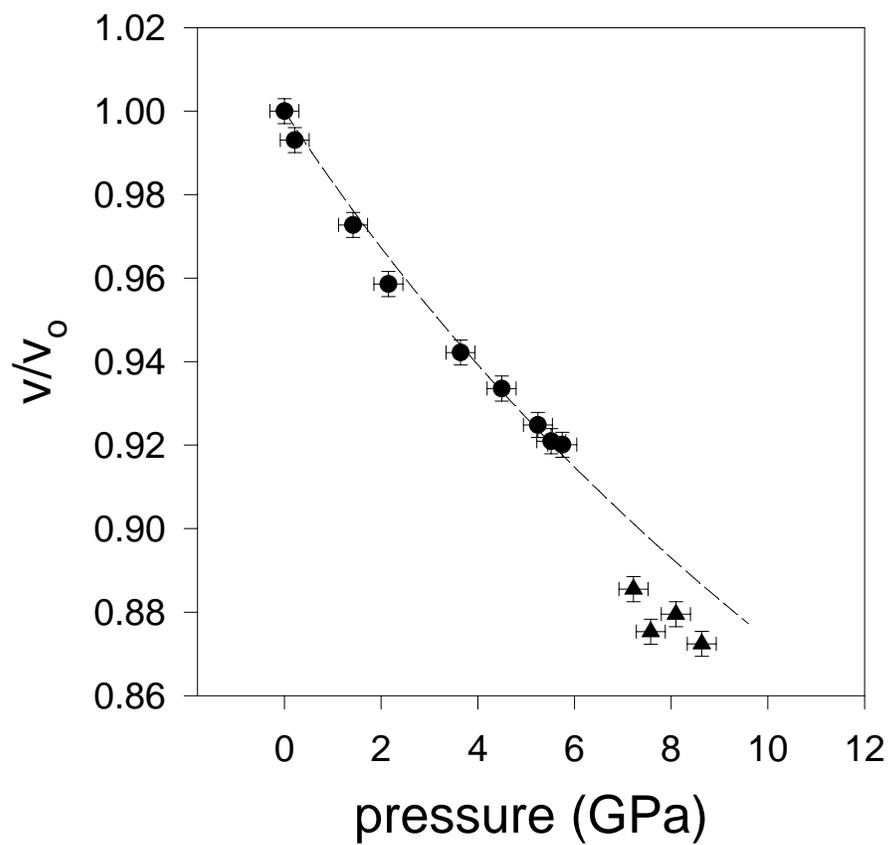

(d) $V/V_0$ of $BaMoO_4$ (●) for the scheelite phase and (▲) for the ferguson­ite phase. In this figure $V_0$ refers to that of the scheelite phase at ambient conditions. The dashed line shows the fit to Birch-Murnaghan equation of state in the scheelite phase.

## Table 1

Observed Raman frequencies, pressure dependence and mode-Gruneisen parameter ($\gamma_i = -d\ln\omega_i/d\ln V = (B_0/\omega_i)(d\omega_i/dp)$ for the scheelite phase of $BaMoO_4$. ($B_0$ is the bulk modulus and $\omega_i$ is the phonon frequency). Column 4 also lists the values of Grüneisen parameters for the modes of the same symmetry in $BaWO_4$ and $CaMoO_4$. For $BaWO_4$, Grüneisen parameter has been scaled to experimental value of the bulk modulus (57 GPa [9]) comapred to the value (40 GPa) used by Jayaraman et al [12]. The observed Raman active modes and their pressure dependence in the fergusonite phase are also listed.

_________________________________________________________________________________

| | Scheelite phase | | | | Fergusonite phase | | |
|---|---|---|---|---|---|---|---|
| $\omega$ (cm$^{-1}$) | $d\omega/dP$ (cm$^{-1}$/GPa) | $\gamma_i$ | $\gamma_i$ BaWO$_4$ (CaMoO$_4$) | Mode Assignment | Raman mode | $\omega_R$ at 6 GPa (cm$^{-1}$) | $d\omega/dP$ (cm$^{-1}$/GPa) |
| 322 | 2.54 | 0.44 | 0.54 (0.82) | $A_g/B_g$ ($\nu_2$) | $\omega 1$ | 309 | -1.34 |
| | | | | | $\omega 2$ | 338 | 0.58 |
| 343 | 3.05 | 0.50 | 0.33 (0.81) | $B_g$ ($\nu_4$) | $\omega 3$ | 351 | 2.89 |
| | | | | | $\omega 4$ | 370 | 4.54 |
| 358 | 3.27 | 0.51 | ---- (0.98) | $E_g$ ($\nu_4$) | $\omega 5$ | 806 | 0.55 |
| | | | | | $\omega 6$ | 816 | 5.08 |
| 792 | 4.25 | 0.30 | 0.23 (0.32) | $E_g$ ($\nu_3$) | $\omega 7$ | 820 | 0.93 |
| | | | | | $\omega 8$ | 835 | 2.49 |
| 839 | 2.87 | 0.19 | 0.16 (0.21) | $B_g$ ($\nu_3$) | $\omega 9$ | 854 | 1.01 |
| | | | | | $\omega 10$ | 866 | 1.77 |
| 892 | 3.66 | 0.23 | 0.17 (0.20) | $A_g$ ($\nu_1$) | $\omega 11$ | 911 | 0.71 |
| | | | | | $\omega 12$ | 923 | 2.84 |

The values of $\omega_3$, $\omega_7$, $\omega_{12}$ have been obtained by extrapolating to 6.0 GPa.

## Table 2

Structural parameters for Scheelite BaMoO$_4$ at ambient condition. The space group is I4$_1$/a (Z = 4) and the lattice parameters are a = 5.58 Å, c = 12.82 Å.

| Atom | Site | x | y | z |
|------|------|-----------|-----------|-----------|
| Ba | 4b | 0.0000 | 0.0000 | 0.5000 |
| Mo | 4a | 0.0000 | 0.0000 | 0.0000 |
| O | 16f | 0.2435(9) | 0.1146(8) | 0.0751(1) |

| | | Bond lengths (Å) | | |
|---|---|---|---|---|
| Mo-O | 1.778 | Ba-O1 | 2.745 | |
| | | Ba-O2 | 2.724 | |

Structural parameter for fergusonite BaMoO$_4$ phase at 7.2 GPa. The space group is I2/a (Z = 4) and the lattice parameters are a = 5.4251 Å, b= 12.3972 Å, c = 5.2599 Å and β = 89.53°.

| Atom | Site | x | y | z |
|------|------|-----------|-----------|-----------|
| Ba | 4e | 0.2500 | 0.6242(4) | 0.0000 |
| Mo | 4e | 0.2500 | 0.1280(8) | 0.0000 |
| O | 8f | 0.8584(2) | 0.9451(6) | 0.2490(2) |
| O | 8f | 0.4973(1) | 0.1998(1) | 0.8836(1) |

| | | Bond lengths (Å) | | |
|---|---|---|---|---|
| Mo-O1 | 1.699 | Ba-O1 | 2.644 | |
| Mo-O2 | 1.713 | Ba-O2 | 2.646 | |
| | | Ba-O3 | 2.598 | |
| | | Ba-O4 | 2.639 | |

Estimated standard deviations are given in parenthesis.